\documentclass[11pt]{article}

\usepackage[letterpaper,margin=1in]{geometry}

\usepackage{fancyhdr}
\usepackage[marginal]{footmisc}

\usepackage{cite}
\usepackage{amsmath,amssymb,amsfonts}
\usepackage{algorithmic}
\usepackage{graphicx}
\usepackage{textcomp}
\usepackage{xcolor}

\usepackage{booktabs}   %% For formal tables:
\usepackage{subfigure}

\usepackage{amsthm}

\usepackage{mathtools}
\usepackage{xcolor}
\usepackage{enumitem}
\usepackage{multirow}
\usepackage{tabularx}
\usepackage{float}

\usepackage{listings}
\usepackage{pifont}

\definecolor{codegreen}{rgb}{0,0.6,0}
\definecolor{codegray}{rgb}{0.5,0.5,0.5}
\definecolor{codepurple}{rgb}{0.58,0,0.82}
\definecolor{backcolour}{rgb}{0.95,0.95,0.92}

\lstdefinestyle{cuda}{
    commentstyle=\color{codegreen},
    keywordstyle=\color{magenta},
    numberstyle=\tiny\color{codegray},
    stringstyle=\color{codepurple},
    basicstyle=\ttfamily\footnotesize,
    breakatwhitespace=false,         
    breaklines=true,                 
    captionpos=b,                    
    keepspaces=true,                 
    numbers=left,                    
    numbersep=5pt,                  
    showspaces=false,                
    showstringspaces=false,
    showtabs=false,                  
    tabsize=2,
    emph={
        constexpr,
        __global__, __shared__, __device__, __host__,
    },
    emphstyle={\color{magenta}},
    frame=lines,
}
\usepackage[
linesnumbered,
commentsnumbered,
lined,
ruled
]{algorithm2e}

\usepackage[hyphens]{url}
\usepackage[unicode,hidelinks]{hyperref}

\usepackage{cleveref}

\newcommand{\rbr}[1]{\left(#1\right)}
\newcommand{\sbr}[1]{\left[#1\right]}

\newcommand{\floor}[1]{\left\lfloor#1\right\rfloor}
\newcommand{\ceil}[1]{\left\lceil#1\right\rceil}
\newcommand{\norm}[1]{\left\vert#1\right\vert}

\newcommand{\dcircle}[1]{\ding{\numexpr201 + #1}}

\newcommand*{\affaddr}[1]{#1} % No op here. Customize it for different styles.
\newcommand*{\affmark}[1][*]{\textsuperscript{#1}}
\newcommand*{\email}[1]{\texttt{#1}}

\title{{RadiK}: Scalable and Optimized {GPU}-Parallel\\Radix Top-K Selection}
\author{%
Yifei Li\affmark[1],
Bole Zhou\affmark[2],
Jiejing Zhang\affmark[1],
Xuechao Wei\affmark[1],
Yinghan Li\affmark[1],
and Yingda Chen\affmark[1]\\
\affaddr{\affmark[1]Alibaba Group},
\affaddr{\affmark[2]Independent}\\
\email{\{lyf383659,jiejing.zjj,yingda.chen\}@alibaba-inc.com}, \email{zhoubole@aliyun.com}
}

\date{}

\begin{document}

\maketitle

\thispagestyle{fancy}
\fancyhf{}
\renewcommand{\headrulewidth}{0pt}
\renewcommand{\footrulewidth}{.5pt}
\fancyfoot[L]{\footnotesize This paper is published at the 38th ACM International Conference on Supercomputing (ICS '24).}

%%%%%% -- PAPER CONTENT STARTS-- %%%%%%%%

\begin{abstract}
Top-k selection, which identifies the largest or smallest $k$ elements from a data set, is a fundamental operation in data-intensive domains such as databases and deep learning, so its scalability and efficiency are critical for these high-performance systems.
However, previous studies on its efficient GPU implementation are mostly merge-based and rely heavily on the fast but size-limited on-chip memory, thereby limiting the scalability with a restricted upper bound on $k$.
This work introduces a scalable and optimized GPU-parallel radix top-k selection that supports significantly larger $k$ values than existing methods without compromising efficiency, regardless of input length and batch size.
Our method incorporates a novel optimization framework tailored for high memory bandwidth and resource utilization, achieving up to 2.5$\times$ speedup over the prior art for non-batch queries and up to 4.8$\times$ speedup for batch queries.
In addition, we propose an adaptive scaling technique that strengthens the robustness, which further provides up to 2.7$\times$ speedup on highly adversarial input distributions.
\end{abstract}

\section{Introduction}
\label{sec:introduction}

Top-k selection, or top-k for short, is a classic algorithmic challenge that involves identifying the $k$ largest or smallest elements out of $n$ input elements according to some predefined sorting criterion.
Top-k is a variant of k-selection that selects exactly the $k$th element.
Often, top-k algorithms return not only the values but also their corresponding indices within the input array.
This is extremely useful not only in traditional scenarios such as high-volume databases~\cite{faiss_2021,root_2016_mapd,mit_topk_2018} where queries may involve selecting the top-$k$ records, but also in cutting-edge deep learning applications such as large language model (LLM) inference services~\cite{fan_2018_neuralstory,holtzman_2020_neuraltext,openai_2019_multitask} where the output is sampled from the most probable $k$ tokens.

In recent years, there has been an explosion in both the amount of data and the amount of computing power required to analyze the data, especially as deep learning has become increasingly important in numerous applications.
The Graphics Processing Unit (GPU) is a well-established type of massively parallel hardware that was originally designed for graphics rendering, but is now becoming a common choice for extreme computing power~\cite{Gokhale_2008_hardware,osti_1494112}.
Thus many popular frameworks or engines for high-volume data processing in various application areas have provided the ability to offload intensive computations to GPUs, such as TensorFlow~\cite{abadi_2016_tensorflow} and PyTorch~\cite{2019_pytorch}, two widely-used deep learning frameworks, MapD~\cite{root_2016_mapd}, a commonly-accepted GPU-powered database, and Faiss~\cite{faiss_2021}, a well-known library for similarity search and clustering of dense vectors.

Under such circumstances, research on top-k sees a trend to migrate from traditional single- or multi-core processors to GPUs in order to make better use of the available computing power and to fit into these frameworks.
A na\"ive idea called ``Sort \& Choose"~\cite{alabi_kselect_2012, 2021_Gaihre_dr_topk, mit_topk_2018} is to sort the entire input and select the $k$ elements from one end of the sorted array.
Although there are fast GPU sorting algorithms such as Radix Sort~\cite{merrill_radixsort_2011, elias_radixsort_2017} and Bitonic Sort~\cite{hagen_bitonicsort_2010, batcher_sort_1968}, this still requires much more work than necessary, especially when $k$ is much smaller than the input length $n$.
There are also attempts to directly adapt the priority queue to GPUs, i.e., maintaining per-thread priority queues to find local top-k, and performing a final reduction across all threads to get the global top-k.
However, per-thread priority queues suffer from thread divergence due to the highly data-dependent thread execution path~\cite{mit_topk_2018}, which should always be avoided in the Single-Instruction Multiple-Thread (SIMT) architecture of GPUs.

For massively parallel hardware like GPUs, merge-based algorithms are often good choices.
There is practical work combining the priority queue and Bitonic Sort to build fast GPU top-k~\cite{faiss_2021}.
Besides, k-selection is also a reasonable starting point for top-k, and as the duality of Bitonic Sort, Bitonic Select ranks a qualified candidate on GPUs~\cite{mit_topk_2018}.
However, the scalability of both these merge-based methods is limited by a restricted upper bound on $k$, typically 512 or 1024, because the $O(k)$ sorting and merging utility must entirely reside in the fast but size-limited on-chip memory.

Such constraints are problematic for real-world applications, such as the k-nearest neighbors (KNN) operation in dense vector databases~\cite{faiss_2021}, where a $k$ up to several thousand is useful in large-scale scenarios, including e-commerce recommendation systems.
Another illustration involves LLM inference engines.
For each token generated by the model, the final sampling step often involves selecting the $k$ tokens with the highest probabilities from the whole vocabulary~\cite{pmlr-2019-beam,pmlr-2023-llm_sampling}.
Both large and small $k$ values have their place~\cite{holtzman_2020_neuraltext}, and real-world LLM services must tolerate $k$ values up to the vocabulary size which can reach 100k for models like GPT-4~\cite{openai_token_2024-1}.
LLM inference engines such as vLLM~\cite{kwon_2023_vllm} commonly adopt ``Sort \& Choose" as a fallback.
However, our performance profiling of a typical inference engine using CUB sort~\cite{nvidia_2022_cub} indicates that sorting alone can consume as much as 28.9\% of the total execution time when serving a model with 7 billion parameters on a single NVIDIA A10 GPU -- a significant bottleneck.

Apart from the merge-based Bitonic Select, distribution-based methods are also promising, such as Radix Select~\cite{alabi_kselect_2012, nvidia_topk_2020} and Bucket Select~\cite{alabi_kselect_2012, AllisonN80_partitioning}, inspired by Radix Sort and Bucket Sort, respectively.
They divide the input into subsets called bins or buckets, and quickly reduce the problem size by focusing only on the bin of the $k$th element, iteration by iteration.
The number of bins is a constant value, so all the bins can reside in on-chip memory of any appropriate size, allowing arbitrary $k\le n$.
However, if $n$ is too large to fit in on-chip memory, the input elements themselves must reside in larger off-chip memory, and it degrades performance to access off-chip memory every iteration.
Moreover, Radix Select is vulnerable to skewed input distributions, and as for Bucket Select, it relies on the predetermined minimum and maximum values of the input which adds overhead.

On the other hand, the need for batch queries must be taken into account, for batching is key to achieving high throughput in real-world applications~\cite{kpp_batch_2021,knk_batch_2022}.
A lack of batch query optimization can result in excessive kernel launch overhead and suboptimal resource utilization.
Additionally, batch queries often involve concatenating multiple input arrays with varying lengths, which may violate address alignment assumptions and consequently reduce memory bandwidth on GPUs.

To implement efficient GPU top-k without compromising scalability and robustness, we introduce a massively parallel radix top-k algorithm that not only supports much greater $k$ than the prior art, but also preserves high efficiency for large $k$ and input lengths, regardless of the batch size.
Based on Radix Select, our method features a carefully designed optimization framework aiming at maximizing memory bandwidth and resource utilization in both batch and non-batch scenarios.
Furthermore, for robustness concerns, we design a scaling technique to overcome the inherent vulnerability of Radix Select with respect to adversarial input distributions.

In summary, we claim the following contributions.
\vspace{-0mm}
\begin{itemize}
    \item We optimize the off-chip memory access for radix top-k with a set of optimizations including hierarchical atomics and flush-efficient write buffer, which empowers our method to outperform the prior art by up to a factor of 2.5 for non-batch queries.
    \item We propose a set of optimizations targeting batch queries, including task rescheduling and offset padding, which further speed up our method by up to a factor of 4.8 for batch queries.
    \item We analyze the impact of adversarial input distributions, propose a lightweight refinement to improve robustness, and demonstrate its effectiveness with experiments.
\end{itemize}
\vspace{-0mm}

The rest of the paper is organized as follows.
\Cref{sec:background} presents the background and our motivation.
\Cref{sec:implementation} details our optimization framework which improves memory bandwidth and resource utilization for higher scalability.
\Cref{sec:discussion} covers a discussion on input data distribution and robustness, as well as our refinement against adversarial distributions.
\Cref{sec:eval} evaluates the performance of this work in various scenarios.
\Cref{sec:related} briefly reviews related work, and finally, \Cref{sec:conclusion} concludes this paper.

\section{Background and Motivation}
\label{sec:background}

\subsection{GPU Programming Model}
\label{sec:background:gpu}

NVIDIA GPUs and the CUDA toolkit~\cite{nvidia_2024_cuda} make one of the common choices for general-purpose GPU programming.
This section briefly introduces some basic concepts of the CUDA GPU programming model~\cite{nvidia_cudaguide_2022} we exploit in our work.

\subsubsection{Thread Hierarchy}
\label{sec:background:gpu:thread}

A function, also known as a \emph{kernel}, is executed in parallel by multiple GPU threads in the Single-Instruction Multiple-Thread (SIMT) model.
A GPU has hundreds of \emph{Streaming Multiprocessors} (SMs), each containing numerous cores.
To map threads to hardware, threads are organized into a hierarchy.
The innermost level called \emph{warp} is a handful of consecutive threads, typically 32, executing the same instruction simultaneously.
Consecutive warps of threads are grouped into a \emph{block}, and all threads in a block execute on the same SM.
Blocks are further grouped into a \emph{grid} and are automatically scheduled by the hardware.

\subsubsection{Memory Hierarchy}
\label{sec:background:gpu:memory}

The highest level of the memory hierarchy is the thread-local \emph{registers}.
CUDA provides \emph{warp primitives} to exchange register data between threads within a warp.
Threads in the same block share a high-speed on-chip \emph{shared memory} (SMEM) of relatively limited capacity, which is useful for inter-warp but intra-block communication.
Registers and SMEM are both SM \textbf{on-chip} resources, while the \textbf{off-chip} VRAM, called \emph{global memory} (GMEM), is shared by all SMs as well as all blocks.
Generally, inter-block communication uses global memory.
Global memory provides much lower bandwidth than shared memory, but its capacity is orders of magnitude larger than on-chip memory.
The on-chip L2 cache helps speed up global memory access.

\subsubsection{Global Memory Access Tricks}
\label{sec:background:gpu:trick}

Several generic optimizations help improve global memory bandwidth.
One trick is called \emph{coalesced memory access pattern}~\cite{Harris_cuda_2013}, where threads within a warp access successive memory addresses simultaneously to improve bandwidth.
Another called \emph{packed data type} helps save the amortized overhead of memory access.
For example, 4 consecutive 8-bit integers can be loaded with only 1 memory load instruction by treating them as a 32-bit integer, as long as the address of the first 8-bit integer is 4-byte aligned.
The maximum size supported by a single memory access instruction on NVIDIA GPUs is 16 bytes.

% ==================================================================================

\subsection{GPU Top-K Algorithms}
\label{sec:background:topk}

\subsubsection{Merge-Based Priority Queue}
\label{sec:background:topk:warpselect}

Although the traditional priority queue implies poor data parallelism, a subtly designed merge-based priority queue, called Warp Select~\cite{faiss_2021}, becomes a state-of-the-art GPU top-k algorithm.
Utilizing a single warp, each thread maintains a sorted thread queue, and all threads together maintain a sorted warp queue whose elements are smaller than those in the thread queues, making the entire data structure a parallel min-heap.
The main idea is to filter inputs with thread queues before updating them to the warp queue with Odd-Size Sort, a merge-based sorting algorithm derived from Bitonic Sort~\cite{hagen_bitonicsort_2010,batcher_sort_1968}.
However, as the input length $n$ increases, it suffers from low parallelism because it employs only one warp to parse the entire input.

Block Select~\cite{nvidia_topk_2020} improves parallelism by organizing 2 -- 4 warps into a block with the help of shared memory to deal with an input array.
It further derives Multi-Block Select, or Grid Select, which coordinates multiple blocks via global memory to further improve parallelism for even larger inputs.
However, it requires tuning as a hyper-parameter the number of elements parsed by each block, to find an optimum between overall parallelism and per-block utilization.
We find this method sensitive to this hyper-parameter in practice, especially when $k$ is large, which hurts usability.

Apart from parallelism concerns, another major drawback is that $k$ is limited by the size of the register file and the SMEM, since the high performance of these merge-based methods relies on keeping the $O(k)$-sized sorting network entirely on chip for high bandwidth.
Both Warp Select and Block Select are implemented as building blocks in Faiss~\cite{faiss_2021} which requires $k$ to be no larger than 1024 or 2048.

% -----------------------------------------------------------------

\subsubsection{Merge-Based Selection}
\label{sec:background:topk:bitonic}

Top-k can be regarded as a variant of k-selection, which is usually the dual of sorting.
Another state-of-the-art GPU top-k algorithms, Bitonic Select~\cite{mit_topk_2018}, is derived from Bitonic Sort~\cite{hagen_bitonicsort_2010, batcher_sort_1968}.
Instead of sorting the entire input, Bitonic Select only needs to generate sorted runs of length $k$, saving a lot of resources.
These sequences are finally merged to produce a single sorted array of length $k$, i.e., the top-k output.
Unfortunately, as a merge-based algorithm, Bitonic Select suffers from the same limitation on $k$ as the merge-based priority queues in \Cref{sec:background:topk:warpselect} due to the on-chip sorting network of size $O(k)$.

% -----------------------------------------------------------------

\subsubsection{Distribution Selection}
\label{sec:background:topk:radix}

Distribution selection algorithms such as Radix Select~\cite{alabi_kselect_2012, nvidia_topk_2020} and Bucket Select~\cite{alabi_kselect_2012, AllisonN80_partitioning} differ from merge-based selection in that they only consume a constant amount of on-chip resources to keep track of a \emph{histogram} that records the number of elements falling into each bin or bucket, allowing them to tolerate a much larger $k$.
Only the bin containing the $k$th element is passed to the next iteration.
In this way, the problem size will be quickly reduced if only a few elements share the same bin with the $k$th element.
Unlike Bitonic Select which outputs all top-k elements, distribution selection only finds the $k$th element, namely the \emph{pivot}.
It is necessary to apply a \emph{filter} to the input array to find all elements greater or less than the pivot~\cite{nvidia_topk_2020}.

The problem is that multiple blocks are required for better parallelism when the input size $n$ is large, in which case the histogram and other intermediate states must be stored in a \emph{workspace} in off-chip global memory for inter-block communication, and this can be troublesome and expensive.
Besides, the performance degrades when the inputs follow skewed distributions, because the target bin may contain a large fraction of the input elements.

% ==================================================================================
\subsection{Motivation and Challenges}
\label{sec:background:motiv}

Our goal is to develop a scalable GPU top-k algorithm that supports sufficiently large $k$ and input length $n$, while maintaining high efficiency in both batch and non-batch scenarios.
We find Radix Select to be a promising candidate, for it has no inherent flaw that prevents $k$ from increasing, unlike the merge-based methods mentioned before.
In this section, we first present an overview of the radix top-k algorithm, and then discuss the challenges we face to achieve our goal.

\subsubsection{Overview of Radix Top-K}
\label{sec:background:motiv:brief}

\begin{figure}[htb]
    \centering
    \includegraphics[width=0.5\linewidth]{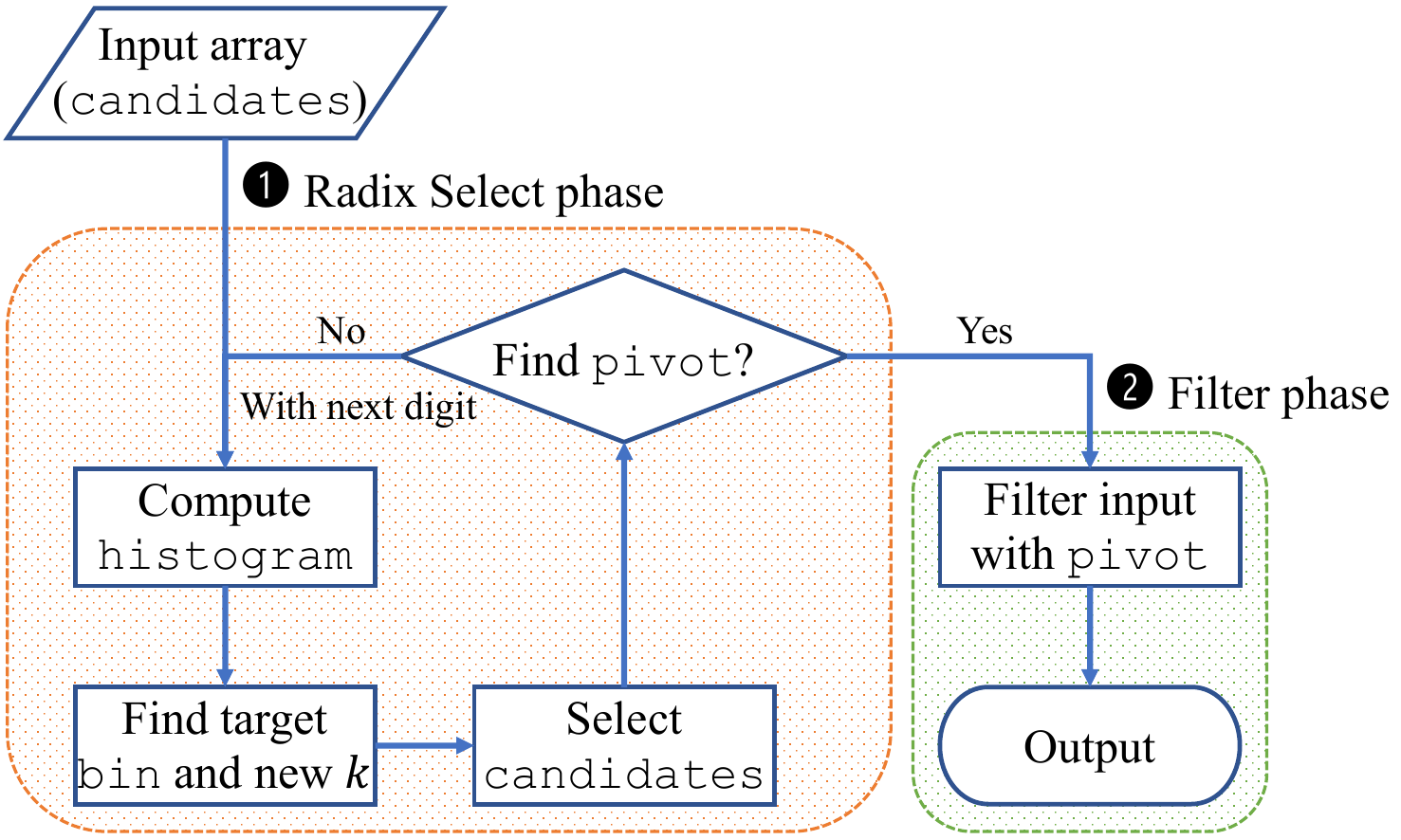}
    \caption{Workflow of radix top-k}
    \label{fig:overview}
\end{figure}

\Cref{alg:topK} presents (GPU) radix top-k with two phases: \dcircle{1} finding the \texttt{pivot}, i.e., the $k$th element, in the \emph{Radix Select phase}; \dcircle{2} filtering out top-k elements from the original inputs with the \texttt{pivot} as a cutoff in the \emph{Filter phase}.
The workflow is shown in \Cref{fig:overview}.

\begin{algorithm}
\small

\SetAlgoLined
\SetKwComment{Comment}{// }{}

\SetKwInOut{Input}{Input}
\SetKwInOut{Output}{Output}

\SetKwData{High}{high}
\SetKwData{Low}{low}
\SetKwData{Histogram}{histogram}
\SetKwData{Bin}{bin}
\SetKwData{Candidates}{candidates}
\SetKwData{Pivot}{pivot}

% \SetKwFunction{Sizeof}{sizeof}
\SetKwFunction{Min}{min}
\SetKwFunction{Max}{max}
\SetKwFunction{CountBin}{countBin}
\SetKwFunction{SelectBin}{selectBin}
\SetKwFunction{SelectCandidate}{selectCandidate}
\SetKwFunction{Filter}{filter}
\SetKwFunction{BitonicSort}{bitonicSort}

\Input{Array $X$ of size $n$, integer $k>0$, digit width $d>0$}
\Output{Array $Y$ of size $k$ containing top-k of $X$}
\Comment{Digit for the target bin is [\Low, \High)}
\High $\gets$ bit width of the element in $X$\;
% \Low $\gets$ \High $-$ \Width\;
\Low $\gets$ \High $-$ $d$\;
\Candidates $\gets$ $X$ \Comment*[r]{all inputs are candidates}
\Comment{Phase 1: Radix Select}
\While{$|$~\Candidates$|$ $>$ $1$ and \High $>$ $0$}{\label{alg:topK:radixselect:start}
    \Histogram $\gets$ \CountBin{\Candidates, \Low, \High}\;
    \Bin, $k$ $\gets$ \SelectBin{\Histogram, $k$}\;
    \Candidates $\gets$ \SelectCandidate{\Candidates, \Bin}\;
    \Comment{Move towards the next digit}
    \High $\gets$ \High $-$ $d$\;
    \Low $\gets$ \Max{\Low $-$ $d$, $0$}\;\label{alg:topK:radixselect:end}
}
\Comment{Phase 2: Filter}
% \Comment{There is only one distinct candidate}
\Pivot $\gets$ \Candidates[0] \Comment*[r]{distinct candidate}\label{alg:topK:pivot}
$Y \gets$ \Filter{$X$, \Pivot, $k$}\;\label{alg:topK:filter}
% \If{outputs are required to be sorted}{
%     $Y$ $\gets$ \BitonicSort{Y}
% }
\caption{Radix top-k for a single input array}\label{alg:topK}
\end{algorithm}

The whole process involves four subroutines:
\begin{itemize}
    \item \texttt{countBin} computes the \texttt{histogram} where each bin records the number of elements that fall in it;
    \item \texttt{selectBin} identifies the \texttt{bin} containing the \texttt{pivot} as well as the new $k$ indicating the rank of the \texttt{pivot} in that \texttt{bin};
    \item \texttt{selectCandidate} selects all elements belonging to the \texttt{bin} from the old \texttt{candidates} to produce new \texttt{candidates};
    \item \texttt{filter} identifies the output top-k elements from the original input with the \texttt{pivot} as a cutoff.
\end{itemize}
Intermediate states such as \texttt{histogram} and \texttt{candidates} are stored in the workspace in global memory.
At the beginning of phase~\dcircle{2}, there must be only one distinct element in the \texttt{candidates}, which is the \texttt{pivot}.
The final output is not sorted by nature, so extra sorting may be required according to the application.

The \emph{digit} width is $d$, and the number of bins equals the number of distinct digits, i.e., the \emph{radix} $r=2^d$.
We assume Most Significant Digit (MSD) Radix Select~\cite{alabi_kselect_2012, elias_radixsort_2017}.
Each pass only focuses on the bin containing the \texttt{pivot}, and only $1 / 2^d$ of the elements are expected to present at the next pass given uniformly distributed inputs.
A larger $d$ implies a faster decrease in the problem size, possibly fewer phase~\dcircle{1} iterations.
The digit width can be extended to 8 bits or even further~\cite{nvidia_topk_2020}.

% -----------------------------------------------------------------
\subsubsection{Challenges of GPU Radix Top-K}
\label{sec:background:motiv:challenge}

Vanilla \Cref{alg:topK} encounters two challenges on GPUs.

\textbf{Global memory access.} 
Among the four subroutines, it is standard procedure~\cite{alabi_kselect_2012, merrill_radixsort_2011, elias_radixsort_2017} for \texttt{selectBin} to find the \texttt{bin} with a parallel prefix scan over the \texttt{histogram}.
However, the other three are more challenging because they write to possibly variable-length arrays in global memory, which require multiple blocks to improve parallelism and may encounter conflicts.
Early literature on Radix Sort suggests allocating many copies of the workspace in global memory to avoid write conflicts~\cite{merrill_radixsort_2011}.
A large-scale reduction is required to merge these copies, and a huge number of global memory accesses can become a bottleneck.
Atomic operations shed new light on this problem~\cite{nvidia_topk_2020}, with which blocks sharing the same global workspace can easily resolve conflicts, eliminating the extra copies of workspace in global memory as well as the number of global memory accesses.
However, a new issue arises in that each thread can issue $O(n)$ atomics, and busy global atomics drastically degrade bandwidth.
To the best of our knowledge, this issue lacks discussion in typical GPU-parallel algorithms, as they neither rely as heavily on atomics as radix top-k nor recognize it as a bottleneck.

\textbf{Batch query elasticity.} \Cref{alg:topK} is not a natural fit for batch queries.
Apart from simply running tasks one by one, another common solution on GPUs to deal with a batch query containing $B$ tasks is to start $B$ folds of thread blocks, one fold per task, and leave the block scheduling to hardware.
However, this does not work as ideally for GPU radix top-k as it does for typical GPU algorithms including the matrix multiplication where the workload keeps steady throughout the whole process.
Instead, the problem size of radix top-k may be greatly reduced in the latter stages of each task, keeping only a small portion of SMs busy, resulting in low overall resource utilization.
Additionally, as mentioned above, radix top-k makes heavy use of global memory access, but concatenated inputs may violate memory alignment assumptions and prohibit some memory access optimizations.

%%%%%%%%%%%%%%%%%%%%%%%%%%%%%%%%%%%%%%%%%%%%%%%%%%%%%%%%%%%%%%%%%%%%%%%%%%%%%%%%%%%
%%%%%%%%%%%%%%%%%%%%%%%%%%%%%%%%%%%%%%%%%%%%%%%%%%%%%%%%%%%%%%%%%%%%%%%%%%%%%%%%%%%
%%%%%%%%%%%%%%%%%%%%%%%%%%%%%%%%%%%%%%%%%%%%%%%%%%%%%%%%%%%%%%%%%%%%%%%%%%%%%%%%%%%
%%%%%%%%%%%%%%%%%%%%%%%%%%%%%%%%%%%%%%%%%%%%%%%%%%%%%%%%%%%%%%%%%%%%%%%%%%%%%%%%%%%
%%%%%%%%%%%%%%%%%%%%%%%%%%%%%%%%%%%%%%%%%%%%%%%%%%%%%%%%%%%%%%%%%%%%%%%%%%%%%%%%%%%
%%%%%%%%%%%%%%%%%%%%%%%%%%%%%%%%%%%%%%%%%%%%%%%%%%%%%%%%%%%%%%%%%%%%%%%%%%%%%%%%%%%

\section{Optimization Framework}
\label{sec:implementation}

To address the challenges above, we design an optimization framework with which we build a scalable GPU radix top-k that achieves a high efficiency for large $k$ and large input length $n$, regardless of batched or non-batched queries.

\begin{figure}[htb]
    \centering
    \includegraphics[width=0.6\linewidth]{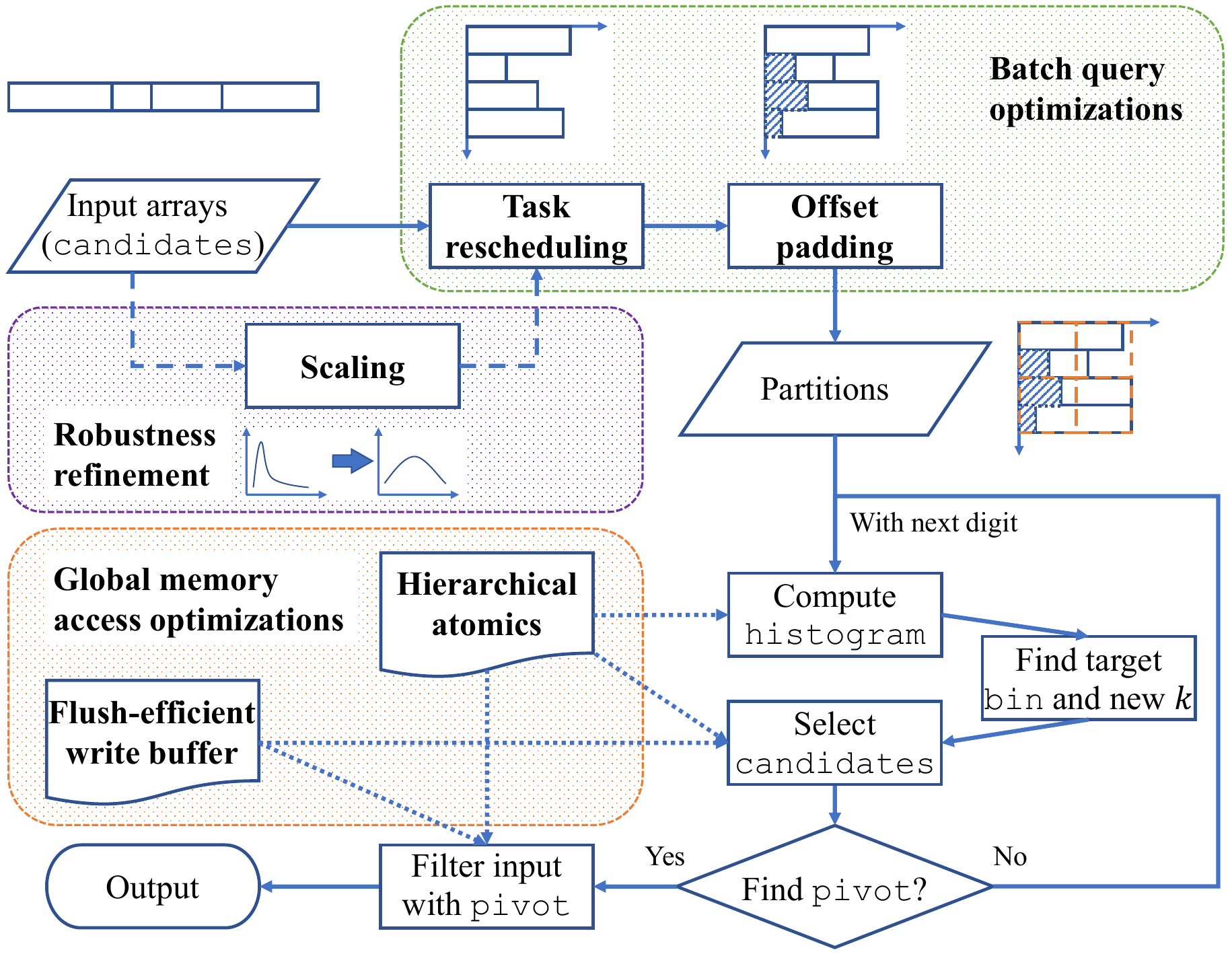}
    \caption{Overview of our optimization framework}
    \label{fig:framework}
\end{figure}

As shown in \Cref{fig:framework}, our framework consists of two sets of optimizations for global memory access and batch queries, respectively.
The optimizations are primarily guided by three common principles in performance optimization~\cite{hennessy_computer_architecture_6ed, nvidia_cudaguide_2022}:
\begin{enumerate}[leftmargin=*]
    \item constructing a hierarchy to balance capacity and speed;
    \item aggregating operations to reduce amortized overhead;
    \item rescheduling and overlapping to hide latency and stalls.
\end{enumerate}
In addition to the two groups of optimizations, we propose a robustness refinement by adaptively scaling the input elements to improve performance given adversarial input distributions.
We highlight our contributions in bold in \Cref{fig:framework}.

% =============================================================================================================

\subsection{Global Memory Access Optimizations}
\label{sec:implementation:memory}

\subsubsection{Hierarchical Atomics}
\label{sec:implementation:memory:atomic}

This optimization is guided by principle (1), built atop the CUDA memory and thread hierarchies.
The discussion below takes \texttt{countBin} as an example, while it also applies to \texttt{selectCandidate} and \texttt{filter}.

\begin{figure}[htb]
\small

\begin{minipage}{1.0\linewidth} % choose width suitably
\begin{lstlisting}[language=c++]
template<typename T, int BLOCK, int LOW, int HIGH>
__global__ void countBin(int *histogram, 
    const T *candidates, int length) {
  constexpr int WIDTH = 1 << (HIGH - LOW);
  // block-local histogram, zero-initialized
  __shared__ int blockHist[WIDTH];
  for (int i = threadIdx.x; i < WIDTH; i += BLOCK)
    blockHist[i] = 0;
  __syncthreads();
  // stage to block-local histogram
  int tid = blockIdx.x * BLOCK + threadIdx.x;
  if (tid < length) {
    int bin = getDigit<LOW, HIGH>(candidates[tid]);
    // this is further optimized to warp primitives
    // and block atomics by compilers
    atomicAdd(blockHist + bin, 1);
  }
  __syncthreads();
  // merge into global histogram
  for (int i = threadIdx.x; i < WIDTH; i += BLOCK)
    if (blockHist[i] > 0)
      // global atomics
      atomicAdd(histogram + i, blockHist[i]);
}
\end{lstlisting}
\end{minipage}
\caption{CUDA pseudo-code for \texttt{countBin} kernel}
\label{lst:countbin-opt}
\end{figure}

The pseudo-code for the optimized \texttt{countBin} is shown in \Cref{lst:countbin-opt}.
Instead of each thread directly maintaining the \texttt{histogram} in the global memory, thread blocks maintain in SMEM their local histograms of the elements assigned to this block.
In this way, the global \texttt{histogram} is distributed to all blocks, and whenever a block finishes, its local ``shard'' is merged into the global \texttt{histogram} with only a constant number of global atomics.
A block-local histogram contains $r=2^d$ bins, regardless of $k$.

Threads within a block still need atomics to update the block-local histogram, but they use cheaper block atomics with fewer side effects.
Further, a warp can aggregate its block atomics with warp primitives\footnote{
The warp primitives are not shown in \Cref{lst:countbin-opt}, because the latest compilers can automatically replace them with warp primitives~\cite{nvidia_warpatomic_2014}.}, reducing the impact of atomics to a greater extent~\cite{nvidia_warpatomic_2014}.
A comparison between different levels of the atomic hierarchy is shown in \Cref{tab:atomic:comp}.

\begin{table}[htb]
    \small
    \centering
    \caption{Levels of the atomic operation hierarchy}
    \label{tab:atomic:comp}
    \begin{tabularx}{\linewidth}{|p{3cm}|X|p{3cm}|p{1.5cm}|}
        \hline
        \textbf{Level of atomics} & \textbf{Executor} & \textbf{Level of memory} & \textbf{Cost} \\\hline
        Warp & All threads within a warp (via warp primitives) & Registers & Very low \\\hline
        Block & One thread from each warp within a block & Shared memory & Low \\\hline
        Grid & One thread from each block within the grid & Global memory & High \\\hline
    \end{tabularx}
\end{table}

% -------------------------------------------------------------------------

\subsubsection{Flush-Efficient Write Buffer}
\label{sec:implementation:memory:writeback}

To reuse thread blocks, when scanning a large array such as the input $X$ or the \texttt{candidates}, the array is divided into \textbf{partitions}.
A block is assigned multiple partitions and processes these partitions one at a time by looping over them.
For simplicity, assume that each thread processes one element out of a partition, then the size of a partition equals the number of threads in a block, denoted by \texttt{BLOCK}.

In \texttt{selectCandidate} and \texttt{filter}, each thread needs to filter elements from partitions and write them back to global memory.
This usually involves a global counter indicating the next position where an element should be written, which needs to be updated with global atomics.
The hierarchical atomics applies to this global counter.
There is another problem: for each element to be written, a thread must perform the write immediately in order to move quickly to the subsequent partitions, resulting in busy and fragmented global memory writes that reduce the memory bandwidth.

In fact, this is similar to the cache system in that whenever a dirty block is evicted, it must be written back to the main memory, but the cache cannot move on until the write is complete, resulting in high latency~\cite{hennessy_computer_architecture_6ed}.
The cache system needs to hide latency, while we need to improve bandwidth, but we can still adopt the write buffer from it.
Elements can be temporarily aggregated in the buffer, and threads perform a burst write when needed, which is more likely to achieve higher memory bandwidth.
Naturally, this buffer resides in SMEM and is shared by all threads within the block.

The remaining problem is to decide the size of this buffer to maximize the efficiency of the buffer flush.
The maximum number of elements a block needs to write to the global memory is equal to the number of elements the block processes, which is proportional to the input size and can be very large.
So we can only reserve a small buffer and flush it periodically before it overflows.
A na\"ive idea is to assign one slot to each thread, and the buffer capacity is \texttt{BLOCK}, which is also the size of a partition.
In this way, after processing each partition, the buffer must be flushed if not empty, because the next partition can overflow the buffer at any time, even if there is only one element in the buffer now.

\begin{figure}[htb]
\small

\begin{minipage}{1.0\linewidth} % choose width suitably
\begin{lstlisting}[language=c++]
template<typename T, int BLOCK, int LOW, int HIGH>
__global__ void selectCandidate(T *candidatesOut, 
    int *globalCount, const T *candidatesIn, 
    int bin, int length) {  
  // buffer capacity equals 2 * BLOCK
  __shared__ T blockCache[2 * BLOCK];
  __shared__ int blockCount[1];
  if (threadIdx.x == 0) blockCount[0] = 0;
  __syncthreads();
  // reuse blocks and loop over partitions
  int tid = blockIdx.x * BLOCK + threadIdx.x;
  int stride = gridDim.x * BLOCK;
  for (int idx = tid; idx < length; idx += stride) {
    T data = candidatesIn[idx];
    // buffer candidate in block shared memory
    if (bin == getDigit<LOW, HIGH>(data)) {
      int pos = atomicAdd(blockCount, 1);
      blockCache[pos] = data;
    }
    __syncthreads();
    // write-back and flush buffer when necessary
    int cnt = blockCount[0];
    __syncthreads();
    if (cnt > BLOCK) {
      if (threadIdx.x == 0)
        blockCount[0] = atomicAdd(globalCount, cnt);
      __syncthreads();
      int base = blockCount[0];
      __syncthreads();
      // aggregated write-back, then reset buffer
      for (int pos = threadIdx.x; pos < cnt; 
           pos += BLOCK)
        candidatesOut[base + pos] = blockCache[pos];
      if (threadIdx.x == 0) blockCount[0] = 0;
      __syncthreads();
    }
  }
}
\end{lstlisting}
\end{minipage}
\caption{Pseudo-code for \texttt{selectCandidate} kernel}
\label{lst:selectCandidate-opt}
\end{figure}

\begin{figure}[htb]
% \small

    \begin{minipage}{0.95\linewidth} % choose width suitably
        \begin{center}
            \includegraphics[width=0.7\linewidth]{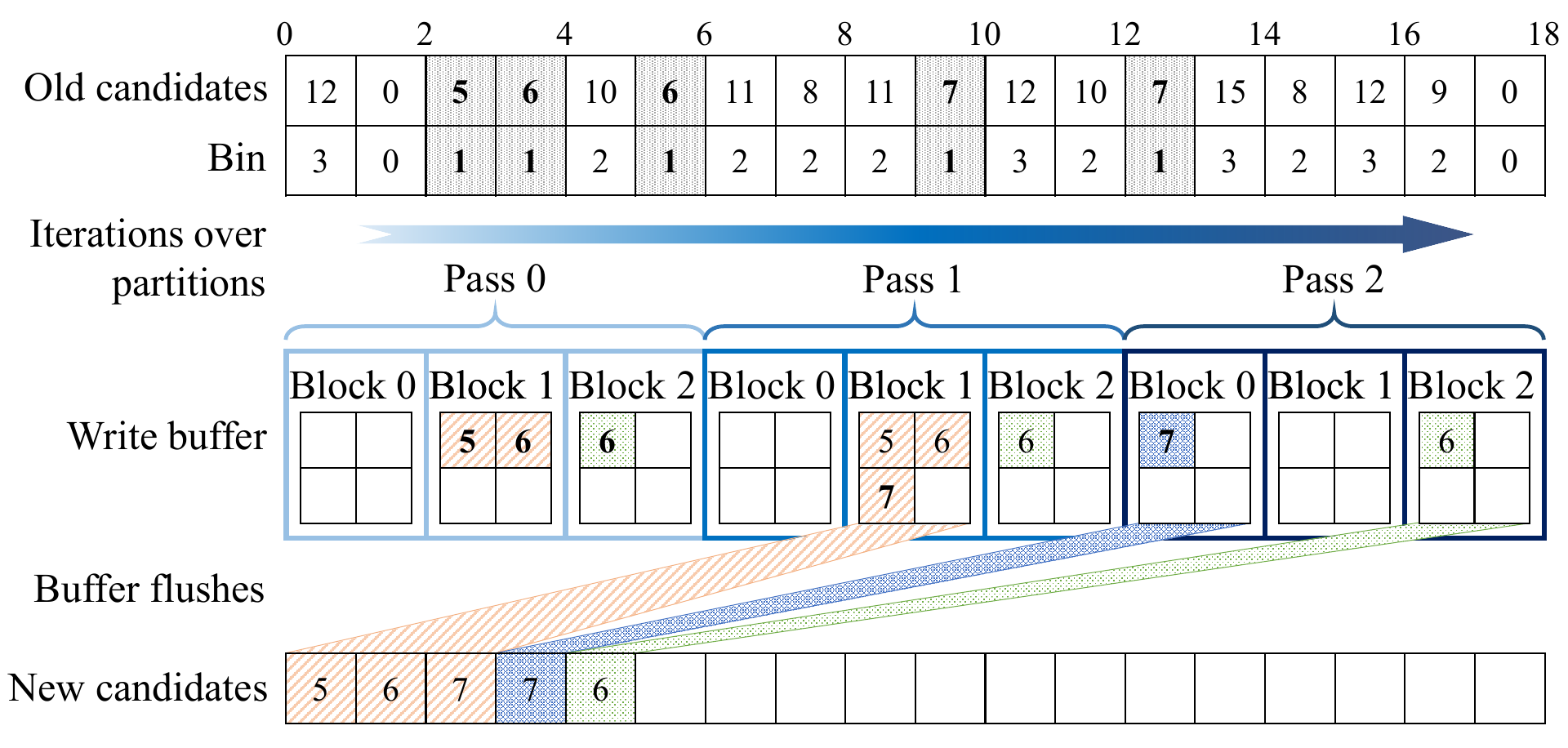}\\
        \end{center}
        {\footnotesize \par Inputs are unsigned 4-bit integers, with $\texttt{BLOCK}=2$, $\texttt{GRID}=3$, $n=18$, $d=2$, and \texttt{bin}=1. The buffer capacity is $2\cdot \texttt{BLOCK} = 4$. The partition size is $2$, and each block processes exactly $\ceil{\frac{n}{\texttt{GRID}\cdot \texttt{BLOCK}}}=3$ partitions, one partition per pass. Block 1 flushes the buffer at the end of pass 1, while a single final flush is sufficient for Block 0 and Block 2.\par}
    \end{minipage}
    \caption{Aggregated write-back with flush-efficient buffer}
    \label{fig:aggregated}
\end{figure}

In \Cref{lst:selectCandidate-opt}, taking \texttt{selectCandidate} as an example, a better solution is to set the buffer capacity to $2 \cdot \texttt{BLOCK}$, and to flush the buffer only when the occupied size exceeds \texttt{BLOCK}, i.e., half of the capacity. 
A block consumes no more than \texttt{BLOCK} slots per pass when looping over partitions.
As long as the occupied size at the end of a pass is less than \texttt{BLOCK}, the empty buffer is sufficient for the next pass.

Let \texttt{GRID} be the grid size, i.e., the number of blocks, and $n$ be the size of the \texttt{candidates}.
Using the buffer of capacity \texttt{BLOCK}, in the worst case, each block flushes every pass, and the number of flushes per block is bounded by
\begin{equation}
\ceil{\frac{n}{\texttt{GRID}\cdot \texttt{BLOCK}}},
\label{eq:no-buf:bound}
\end{equation}
i.e., the number of partitions assigned to each block.
More precisely, suppose the inputs are uniformly distributed, i.e., an element falls into the target \texttt{bin} with probability $1/{2^d}$.
Then the \textbf{expected} number of flushes per block \textbf{per pass} is
\begin{equation}
1 - \rbr{\frac{2^d-1}{2^d}}^{\texttt{BLOCK}}.
\label{eq:no-buf:per-pass}
\end{equation}
Thus the \textbf{expected} number of \textbf{all} flushes per block is
\begin{equation}
\sbr{1 - \rbr{\frac{2^d-1}{2^d}}^{\texttt{BLOCK}}}\ceil{\frac{n}{\texttt{GRID}\cdot \texttt{BLOCK}}}.
\label{eq:no-buf:all}
\end{equation}
For large input length $n$, \texttt{BLOCK} is typically selected to be sufficiently large, and the value of \cref{eq:no-buf:per-pass} is close to $1$ (e.g., when $\texttt{BLOCK}=1024$ and $d=8$, this value is about 0.98), which implies that \cref{eq:no-buf:bound} is a tight upper bound of \cref{eq:no-buf:all}.

On the contrary, with the flush-efficient buffer of capacity $2 \cdot \texttt{BLOCK}$, each flush at least aggregates $\texttt{BLOCK} + 1$ elements.
Given uniformly distributed inputs, $1/2^d$ of the \texttt{BLOCK} elements parsed by each block per pass are expected to fall into the target \texttt{bin}.
Then the \textbf{expected} number of \textbf{all} flushes per block is bounded by
\begin{align}
    &\frac{\texttt{BLOCK}}{2^d(\texttt{BLOCK} + 1)}\ceil{\frac{n}{\texttt{GRID}\cdot \texttt{BLOCK}}}\\
    \approx&\frac{1}{2^d}\ceil{\frac{n}{\texttt{GRID}\cdot \texttt{BLOCK}}},
\end{align}
which is reduced by about a factor of $2^d$ compared with \cref{eq:no-buf:bound}, while the buffer capacity increases only by a factor of $2$.

\Cref{fig:aggregated} gives an example of aggregated write-back to global memory where only one flush per block is sufficient.
Such aggregated write-back not only obeys principle (2), but also fits well with the coalesced memory access pattern and the packed data type, further improving memory bandwidth.

% -------------------------------------------------------------------------

Under some circumstances, the number of flushes can be further reduced.
The \texttt{filter} kernel is very similar to the \texttt{selectCandidate} kernel, except that \texttt{filter} returns a fixed-length array of exactly $k$ elements.%, providing extra optimization opportunities.
Taking advantage of the template metaprogramming, we design a series of specialized \texttt{filter} kernels targeting different sizes of $k$.
Using the write buffer capacity as a template argument, an example of such specialization can be kernels with buffer capacities of $128$, $256$, and so on.
If $k\le 128$, the \texttt{filter} with buffer capacity $128$ will be used.
For very large $k$, we also define a fallback kernel with the reusable buffer strategy described earlier.
The advantage of a fixed-capacity buffer is that it needs to be flushed only once.

% -------------------------------------------------------------------------

\subsection{Batch Query Optimizations}
\label{sec:implementation:batch}

\subsubsection{Task Rescheduling}
\label{sec:implementation:batch:reschedule}

The optimizations in \Cref{sec:implementation:memory} are sufficient for the non-batch query with only one input array. 
However, as mentioned in \Cref{sec:background:motiv:challenge}, due to the particular temporal dynamics of the radix top-k problem size, the goal of efficient batch queries requires more effort to improve the resource utilization in the latter stages of the Radix Select phase.

\begin{figure}[ht]
    \centering
    \subfigure[before]{
        \centering
        \includegraphics[height=4cm]{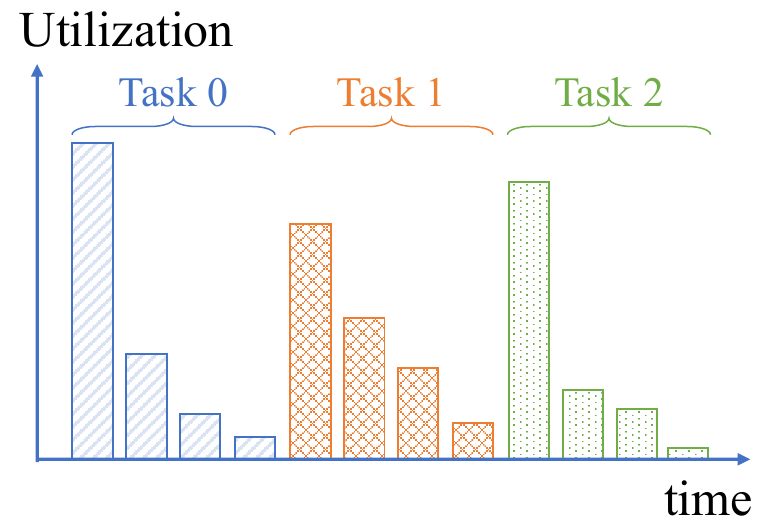}
        \label{fig:reschedule:before}
    }
    % \hfill
    \subfigure[after]{
        \centering
        \includegraphics[height=4cm]{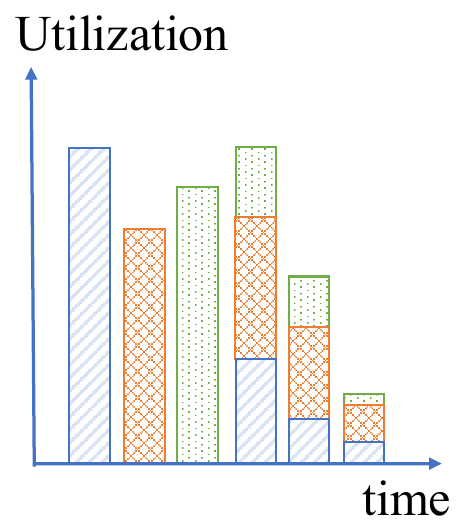}
        \label{fig:reschedule:after}
    }
    \caption{Task rescheduling improves utilization}
\end{figure}

As shown in \Cref{fig:reschedule:before}, for each task in a batch, the first iteration usually achieves a high utilization, while the following iterations require only a relatively small fraction of the computing power.
Based on this observation and inspired by principle (3), we reschedule tasks as in \Cref{fig:reschedule:after}, grouping the first iterations of all tasks horizontally, and stacking each one of the subsequent iterations of all tasks vertically, i.e., the second iteration of all tasks running in parallel, and so on.
In this way, following principle (2), we can further reuse thread blocks between stacked tasks by dividing partitions in both the input length dimension and the task index dimension, reducing the cost of kernel launches.

\subsubsection{Offset Padding on Demand}
\label{sec:implementation:batch:align}

\begin{figure}[ht]
% \small
    \centering
    \begin{minipage}{0.95\linewidth} % choose width suitably
        \begin{center}
            \includegraphics[width=0.8\linewidth]{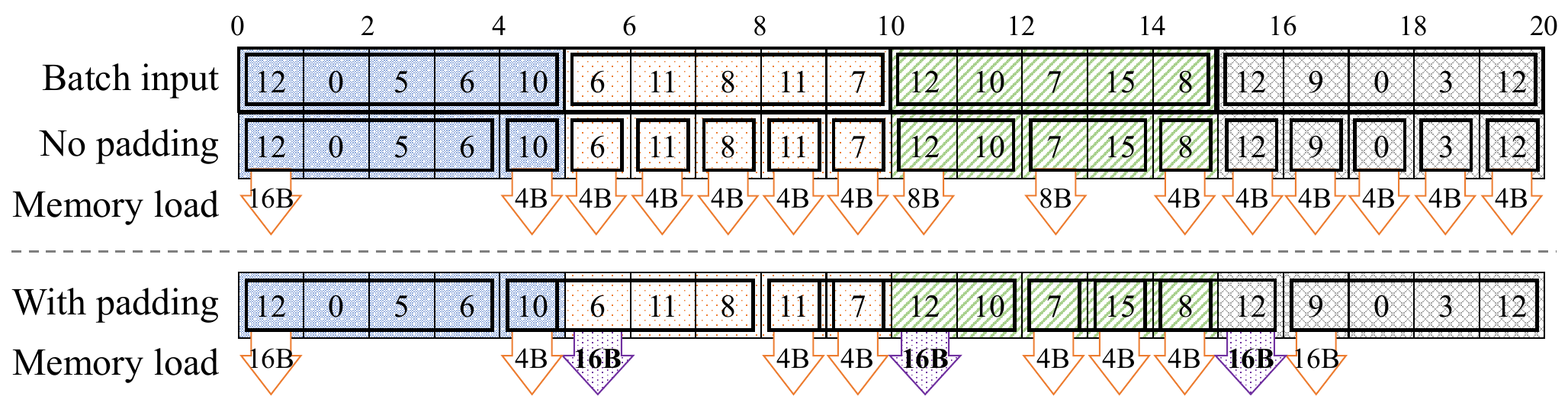}
        \end{center}
        {\footnotesize \par Four arrays of 32-bit integers are concatenated into a compact batch input, each containing 5 elements. Without padding, 15 global memory load transactions are required due to alignment restrictions. With on-demand offset padding, however, only 11 transactions are sufficient. The packed data type trick applies to every task and \texttt{PACKSIZE} is $16$ bytes here. Highlighted 16-byte memory loads contain leading padding. \par}
    \end{minipage}
    \caption{On-demand padding saves memory transactions}
    \label{fig:padding}
\end{figure}

To save bandwidth, the input arrays of a batch query are often concatenated into a compact stride-free array, together with a much smaller array containing the start addresses, i.e., offsets, of each input.
However, the length of each task varies and the offsets do not always align to some convenient bound for the packed data type, preventing the use of this trick.
We propose a simple fix by left-padding input array offsets on demand.
If the size of the packed data type is \texttt{PACKSIZE}, we use the nearest \texttt{PACKSIZE}-aligned address to the left of the original start address as a surrogate.
When loading packed elements from the global memory, the surrogate offset is used instead of the original offset.
Only the first pack is aware of the padding, so we simply patch the thread whose global thread index on the input length dimension equals $0$ to omit the leading padding.
An example is given in \Cref{fig:padding}.

\section{Discussion on Distribution and Robustness}
\label{sec:discussion}

The optimizations in \Cref{sec:implementation} have ensured scalability and efficiency in most cases, but the vulnerability with respect to adversarial input distributions is still a nonnegligible concern.
In this section, we will first discuss the input data distribution in real-world applications.
Then, we will analyze the impact of adversarial distributions and propose an adaptive scaling technique to improve the robustness.

\subsection{Data Distribution in Applications}
\label{sec:discussion:distribution}

Two of the important applications of top-k include the vector database and the LLM inference.
For the vector database, as mentioned in \Cref{sec:introduction}, a common scenario is the KNN query.
As a preprocessing step, the base vectors are often clustered around several centroids.
During the KNN query service, top-k parses the distances between pairs of multidimensional vectors in the same k-means cluster~\cite{faiss_2021,radu_2020_clustering}, and researchers have shown that these distances follow approximately normal distributions~\cite{philip_2007_box,thirey_2015_nspace}, implying low skewness.

For the LLM inference, top-k handles the discrete token sampling probability which is usually the softmax of model output scores over a vocabulary whose length is 100k to 150k~\cite{openai_token_2024-1,bai_2023_qwen}.
The distribution of the probability is rather diverse~\cite{holtzman_2020_neuraltext}, where the so-called \emph{flat distribution} is not rare.
The flat distribution occurs when a number of tokens are similarly suitable for the next word, resulting in a distribution of low skewness.
On the contrary, in the \emph{peaked distribution} one or two tokens concentrate most of the probability mass, e.g., 0,8 or even 0.9, suggesting a nearly deterministic choice of the next token to be generated, while the probabilities of the other tokens are almost uniformly distributed within a narrow range.
Soon we will show that the peaked distribution can be adversarial for radix top-k.

\subsection{Robustness Analysis}
\label{sec:discussion:robust}

Radix-based methods have been known to be vulnerable to skewed distributions since their proposal.
Treating inputs as bit strings, MSD radix top-k scans from the most significant end to the other with the digit width $d$ as step size, and breaks whenever only one candidate corresponds to the digit to scan.
Essentially, the speed of this algorithm depends on the number of Radix Select iterations.
If a large fraction of the inputs cluster near the pivot, i.e., the $k$th elements, many elements will share the same highest digits with it.
Under such circumstances, in the early iterations, these elements will fall into the same bin as the pivot, preventing rapid reduction of the problem size.

\begin{figure}[ht]
    \centering
    \includegraphics[width=0.6\linewidth]{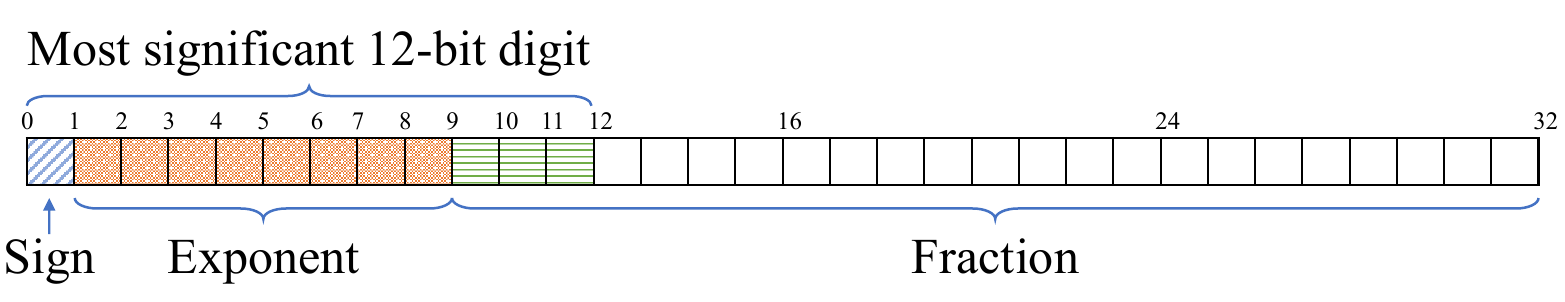}
    \caption{Most significant 12-bit digit of FP32}
    \label{fig:scaling}
\end{figure}

GPUs are mostly used for floating-point arithmetic.
To analyze the robustness of radix top-k, we take the 32-bit floating point (FP32) as an example.
IEEE-754 floating point, the most common layout of FP32, defines the most significant bit as the \emph{sign} bit, the next 8 bits as the \emph{exponent}, and the lower 23 bits as the \emph{fraction}.
We choose a 12-bit digit, so in the first pass, the digit contains the sign bit, all of the exponent bits, and the highest 3 fraction bits, as shown in \Cref{fig:scaling}.

Consider an extreme case where all $n$ input elements $a_0, a_1, \dots, a_n$ fall into the same bin after the first pass, i.e., they share the same highest 12 bits.
Then it satisfies that:
\begin{enumerate}[leftmargin=*]
    \item they are of the same sign;
    \item they are of the same exponent, denoted $h$, and we have $h=\floor{\log_2 \norm{a_i}} $ for every $a_i$;
    \item for any two different keys $a_i$ and $a_j$, we have
    \begin{equation}
        \norm{a_i - a_j} < 2^{h-3} = 2^{\floor{\log_2 \norm{a_i}} - 3},
        \label{eq:scale:origin}
    \end{equation}
    because they share the highest 3 fraction bits representing $2^{h-1}$, $2^{h-2}$, and $2^{h-3}$, respectively.
\end{enumerate}

This reveals the fact that the performance of radix top-k is NOT solely determined by the skewness of the input distribution.
Instead, the choice of the digit width $d$, the data format, and the range of the inputs all play significant roles.
For example, whatever the distribution, if the range of the FP32 inputs is $[128.6,128.7]$, then all the inputs will share the same sign bit which is 0, the same exponent which is 7 (encoded as 0x86), and the same highest 3 fraction bits which are all 0.
Given $d=12$, it is easy to see that in the first iteration, all the inputs fall into the same bin, resulting in a highly adversarial case, even if the inputs simply follow a uniform distribution.

This suggests that the peaked distribution in the LLM inference can be adversarial.
As long as $k$ is larger than one or two, i.e., the pivot does not belong to the most outstanding tokens, a lot of tokens may share the same bin with the pivot in the first iteration and degrade the performance.

\subsection{Adaptive Scaling Technique}
\label{sec:discussion:scaling}

Now that the source of the vulnerability is identified, we propose a lightweight and adaptive refinement for robustness by randomly drawing an element $a_s$ from the input elements and then subtracting it from all elements, which adds little overhead.
This technique is a kind of \emph{scaling}, since floating points will adjust their exponents accordingly.
After scaling, the upper bound of the distance between the adjusted $a_i$ and $a_j$ remains unchanged:
\begin{equation}
    \norm{(a_i-a_s) - (a_j-a_s)}=\norm{a_i - a_j} < 2^{\floor{\log_2 \norm{a_i}} - 3}.
\end{equation}
But the threshold that $a_i-a_s$ and $a_j-a_s$ still cannot be distinguished in the first iteration now becomes
\begin{equation}
    \norm{(a_i-a_s) - (a_j-a_s)} < 2^{\floor{\log_2\norm{a_i-a_s}} - 3}.
\end{equation}
As long as $\norm{a_i} > \norm{a_i-a_s}$, it holds that
\begin{equation}
    2^{\floor{\log_2 \norm{a_i}} - 3} > 2^{\floor{\log_2\norm{a_i-a_s}} - 3},
\end{equation}
which means that the adjusted $a_i$ and $a_j$ are more likely to be distinguished in the first iteration of Radix Select.

We justify the choice of $a_s$ as follows.
If the adversarial distribution is unimodal, the condition $\norm{a_i} > \norm{a_i-a_s}$ can be easily satisfied.
If it is bimodal, suppose one of the peaks is near the $k$th element, namely peak A, and the other peak B.
The more elements cluster around peak A, the more the performance of radix top-k should have degraded, but the more likely it is that the $a_s$ is near peak A, thus satisfying $\norm{a_i} > \norm{a_i-a_s}$ for that peak.
Even if $a_s$ is chosen near peak B, at least peak B will be scattered in this iteration, and peak A will be most likely to be scattered in the next iteration.
As the number of peaks continues to increase, the adversarial impact decreases inversely.

\section{Evaluation}
\label{sec:eval}

\subsection{Methodology}
\label{sec:eval:meth}

To better illustrate the advantages of our top-k query algorithm, we conduct the following experiments:
\begin{enumerate}[leftmargin=*]
    \item varying $k$ and input length for non-batch queries;
    \item varying $k$, input length, and batch size for batch queries;
    \item ablation study on optimizations;
    \item cases where $k$ is up to a quantile;
    \item evaluating scaling technique on adversarial distributions.
\end{enumerate}
Experiments 1 -- 3 evaluate our optimizations, experiment 4 evaluates the scalability when $k=O(n)$, and experiment 5 evaluates our refinement on robustness.
These experiments show that our algorithm remains efficient for very large $k$ and input length $n$, for both non-batch and batch queries, and for either adversarial or non-adversarial distributions, proving its efficiency, scalability, and robustness.

\subsection{Setup}
\label{sec:eval:setup}

We compare this work with the following state-of-the-art methods mentioned in \Cref{sec:background:topk:bitonic,sec:background:topk:warpselect}.

\begin{itemize}
    \item Bitonic Select (\textbf{bitonic})~\cite{mit_topk_2018}: we use the open-source codes by the original authors\footnote{\url{https://github.com/anilshanbhag/gpu-topk}}.
    \item Priority queue with Block Select (\textbf{PQ-block})~\cite{faiss_2021}: we use the open-source codes by Faiss\footnote{\url{https://github.com/facebookresearch/faiss}}.
    \item Priority queue with Grid Select (\textbf{PQ-grid})~\cite{nvidia_topk_2020}: we implement it by ourselves.
\end{itemize}

Test programs are built with CUDA toolkit version 11.6. We run the experiments on a server running \textit{CentOS 7.4} with kernel version 4.9, equipped with the following hardware:
\begin{itemize}
    \item CPU: 2 $\times$ \textit{Intel Xeon Platinum 8163} @ 2.50 GHz (\textit{Skylake} with 24 cores, 48 threads);
    \item GPU: \textit{NVIDIA Tesla T4} @ 1.59 GHz with VRAM @ 5.00 GHz (\textit{Turing}, driver version 525.60.13).
\end{itemize}

We use the FP32 data type in our evaluation.
The results should not show substantial differences for other common data types such as 16-bit floating point (FP16) and brain floating point (bfloat16 or BF16), because our radix-based method treats any data type as a bit string.
Theoretically, our method should perform even better with these 16-bit types which are commonly used in the latest deep learning applications, since a shorter type implies fewer iterations.

Some hyper-parameters such as the digit width $d$ and the thread block size \texttt{BLOCK} are mentioned in our work.
In real-life products, they should be tuned to meet business requirements and deployed hardware, while in our evaluation, we use a set of parameters showing good performance in representative test cases.
Specifically, we choose $d=12$ of which our hardware is capable.
As for \texttt{BLOCK}, in general, it should be adjusted according to the input size.
It should be noted that our method shows a stable performance with respect to reasonable choices of hyper-parameters.

For a fair comparison, if not specified, outputs will be sorted and all results will include the sorting penalty in the following evaluation.
In most cases, this cost only takes a negligible portion, e.g., 0.1\% to 0.6\% in \Cref{fig:exp:multibatch-2}, of the total execution time of our method.

\subsection{Optimization Results}
\label{sec:eval:opt}

\subsubsection{Non-Batch Query}
\label{sec:eval:opt:basic}

To compare the performance of our proposed method with the state of the art in non-batch scenarios, we perform experiments on different combinations of $k$ and input length $n$.
The inputs are randomly sampled from the uniform distribution over $[0, 1]$.
As mentioned in \Cref{sec:background:topk:bitonic}, Bitonic Select only supports $k\le 512$.
The latest open-source implementation of PQ-block supports $k \le 2048$ on our platform.
We do not evaluate PQ-grid for $k \ge 1024$, yet its performance has already dropped to a rather low level when $k = 512$.

\begin{figure*}[htb]
    \centering
    \includegraphics[width=\textwidth]{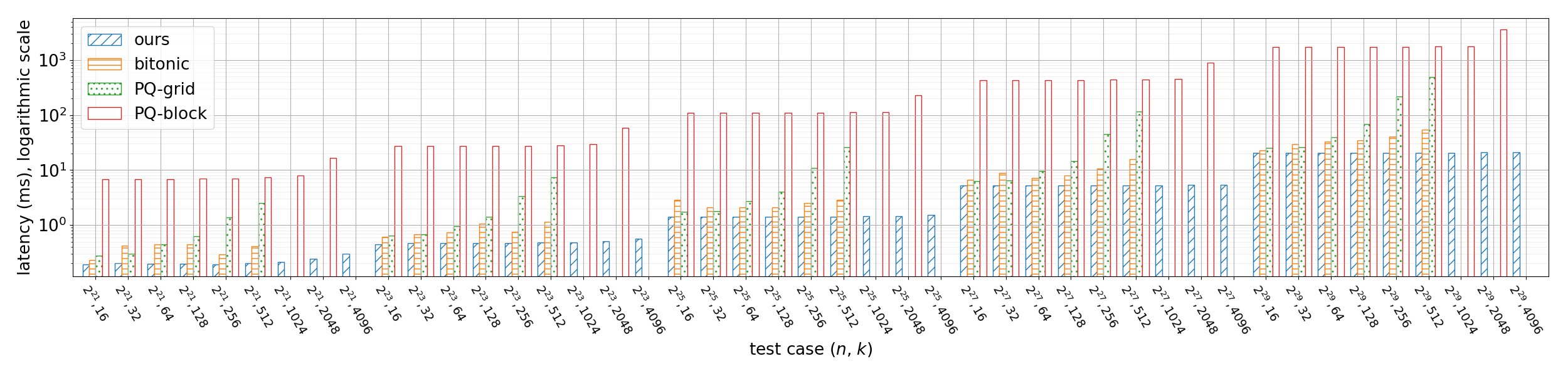}
    \caption{Latency (logarithmic scale) in ms for non-batch query, $k$ and $n$ are powers of $2$, $k$ from $16$ to $4096$, $n$ from $2^{21}$ to $2^{29}$}
    \label{fig:exp:single-query}
\end{figure*}

\begin{figure}[htb]
    \centering
    \includegraphics[width=0.6\linewidth]{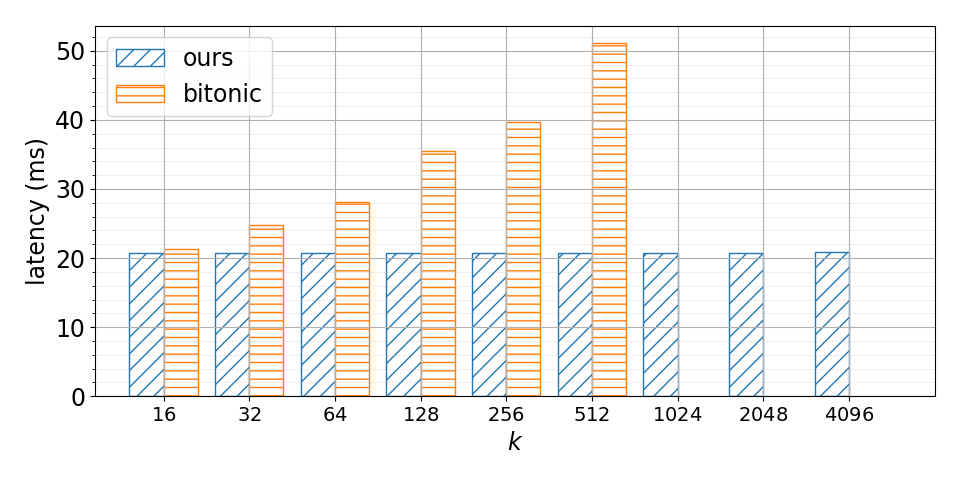}
    \caption{Latency in ms for non-batch query, $n=2^{29}$}
    \label{fig:exp:single-query:part}
\end{figure}

The results are shown in \Cref{fig:exp:single-query}, where the y-axis is in logarithmic scale because PQ-block is orders of magnitude slower than the others.
Our method outperforms all the others in terms of both efficiency and scalability, being the fastest in all test cases, and the only one capable of $k>2048$.
It gains increasing advantage as $k$ and $n$ getting larger, and as shown in \Cref{fig:exp:single-query:part}, when $k=512$ and $n=2^{29}$, our method achieves up to 2.5$\times$ speedup over Bitonic Select, the second fastest method in nearly all cases.
Moreover, the performance of this work remains incredibly stable as $k$ increases, while the latency of the other methods increases rapidly.
PQ-grid shows comparable performance only when $k\le 128$.
PQ-block performs disappointingly for any $n\ge 2^{21}$ because its parallelism is too low for such large inputs.
For comparison, the evaluation of this scenario is limited to inputs no larger than $2^{16}$ in the original paper~\cite{faiss_2021}.

% ----------------------------------------------------------------

\subsubsection{Batch Query}
\label{sec:eval:opt:multibatch}

In the batch query evaluation, we omit PQ-grid because its latency is too much larger than the other methods, though it performs better than PQ-block in non-batch queries.
This is mainly due to the choice of its hyper-parameter which requires careful tuning for any relatively large batch size.
PQ-block supports batch queries by nature, while Bitonic Select does not, so we simply loop it over tasks.
Although in practice the length of each task may be different in a batch query, in this section we use isometric tasks in every batch, because PQ-block requires so.
This cancels the benefit of offset padding, but later we will measure its effect with an ablation study in \Cref{sec:eval:opt:ablation}.
We arrange three sets of experiments, each varying one of $k$, the input length $n$ of each task, and the batch size, with the other two factors fixed.
The inputs follow the uniform distribution on $[0, 1]$.

\begin{figure}[htb]
    \centering
    \subfigure[batch size $=16$, $k=512$]{
        \centering
        \includegraphics[width=0.4\linewidth]{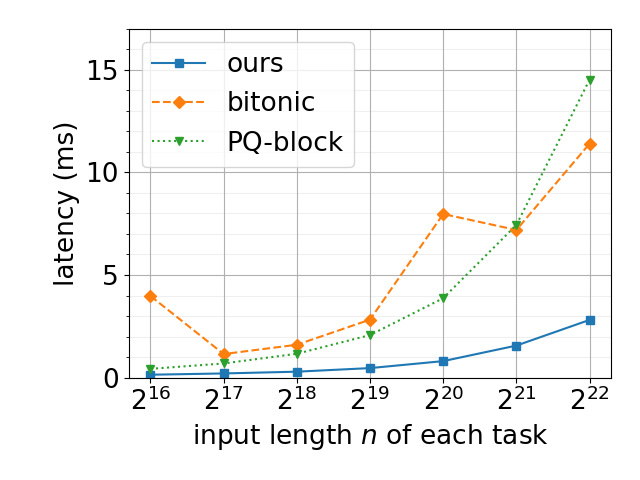}
        \label{fig:exp:multibatch-1}
    }
    \subfigure[batch size $=16$, $n=2^{22}$]{
        \centering
        \includegraphics[width=0.4\linewidth]{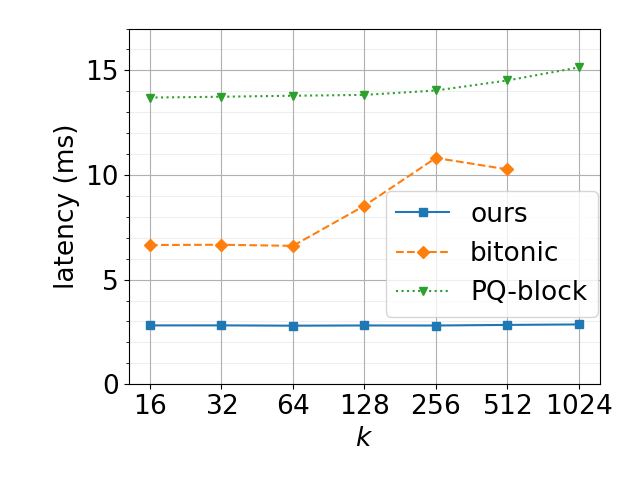}
        \label{fig:exp:multibatch-2}
    }
    \\
    \subfigure[$n=2^{22}$, $k=512$]{
        \centering
        \includegraphics[width=0.4\linewidth]{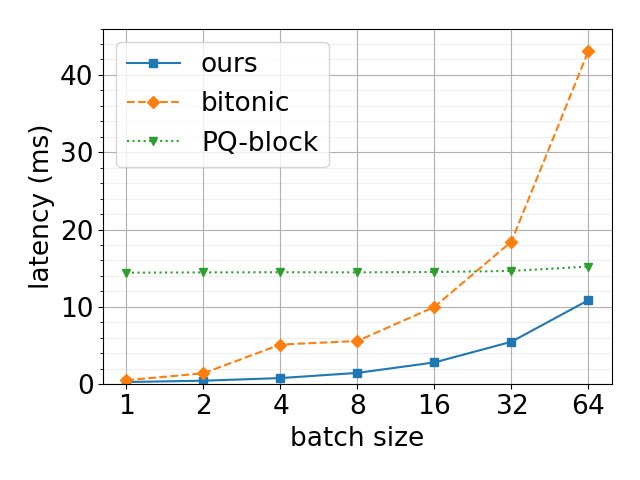}
        \label{fig:exp:multibatch-3}
    }
    \caption{Latency in ms for batch query}
    \label{fig:exp:multibatch}
\end{figure}

The results are shown in \Cref{fig:exp:multibatch}.
Our method still performs the best in all test cases, achieving up to 4.8$\times$ speedup over PQ-block when $n=2^{20}$ in \Cref{fig:exp:multibatch-1}.
Its latency also presents a satisfactory increasing rate as $n$ and $k$ increase.

In \Cref{fig:exp:multibatch-3}, our method outperforms the others though, its latency increases faster than that of PQ-block.
The results suggest that our proposed method is best at the use cases where $n$ is large and the batch size stays in a reasonable range.
We justify this scenario by the fact that in the latest practical LLMs, the batch size almost always remains no larger than 8 or 16 due to the per-task memory pre-allocation~\cite{osdi2022_orca}, while $n$ is often the vocabulary size which has reached 100k (nearly $2^{17}$) for GPT-4 and keeps growing~\cite{openai_token_2024-1}.

% ----------------------------------------------------------------

\subsubsection{Ablation Study}
\label{sec:eval:opt:ablation}

To measure the contribution of each optimization, we design an ablation study.
Since it is hard to completely decouple the hierarchical atomics from the write buffer, we regard them as a single component labeled (1).
The other two components are: (2) task rescheduling, and (3) offset padding.
The baseline is a na\"ive implementation following \Cref{alg:topK}.
For each component, we evaluate its isolated variant which only applies this component to the baseline, and its exclude-one variant which applies all the other components to the baseline.
The full variant includes all three components.
We choose a representative scenario where the batch size is $16$ and $k=2048$.
To present the effect of (3), we set the length of the first task to $n=2^{20}-1$, and use $n=2^{20}$ for the other tasks, so that their offsets are all odd and thus prohibit packed data type.
The inputs follow the uniform distribution on $[0, 1]$.

\begin{figure}[htb]
    \centering
    \begin{minipage}{0.95\linewidth} % choose width suitably
        \begin{center}
            \includegraphics[width=0.6\linewidth]{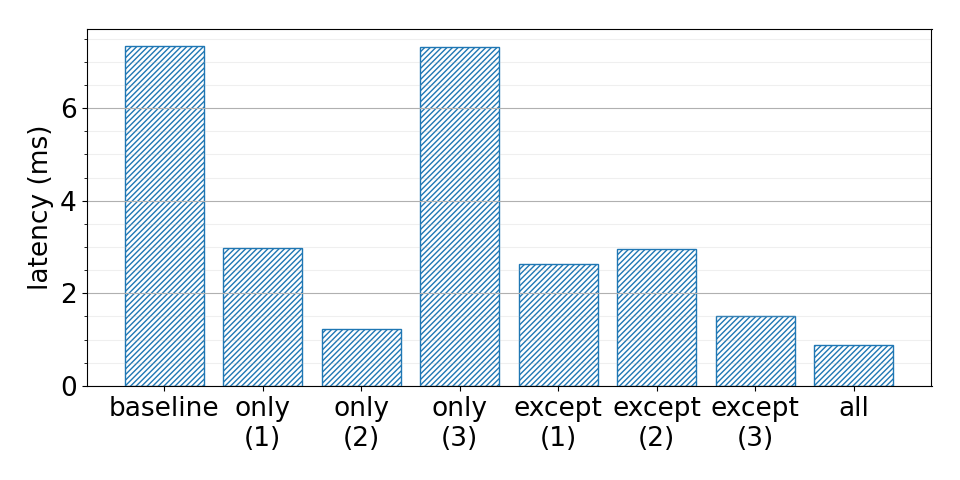}\\
        \end{center}
        {\footnotesize \par The columns labeled ``only'' are the isolated variants, while the columns labeled ``except'' are the exclude-one variants.\par}
    \end{minipage}
    \caption{Ablation study results}
    \label{fig:exp:ablation}
\end{figure}

The results are shown in \Cref{fig:exp:ablation}.
Components (1) and (2) each considerably decrease the latency, while (3) has no effect by itself, because without (2), each task is regarded as a single non-batch query, which prevents the packed data type in order to avoid invalid memory address.
However, it is counterintuitive that except-(1) and except-(3) are both worse than only-(2).
Note that (1) improves global memory write but adds per-element overhead, while (3) improves global memory read but decreases the parallelism because each thread parses more elements.
We can infer from the phenomenon that (2) provides the major part of the speedup in batch queries and the negative effect of (1) or (3) cancels its performance gain. 
Combining all components together, however, with more elements loaded by (3), the longer per-element computation pipeline required by (1) can overlap to hind latency. 
Thus the full variant shows the best performance.

% ------------------------------------------------

\subsection{Quantile Results}
\label{sec:eval:quantile}

To better illustrate the scalability of our proposed method with respect to $k$, we further evaluate three cases where $k=O(n)$, including the percentile, the quartile, and the median.
The batch size and the input length $n$ are arranged the same as \Cref{fig:exp:multibatch-1}.
Since the prior art cannot support such large $k$, we simply compare the performance of our method in different settings, using a constant $k=512$ as the baseline which is also the case in \Cref{fig:exp:multibatch-1}.

\begin{figure}[htb]
    \centering
    \includegraphics[width=0.4\linewidth]{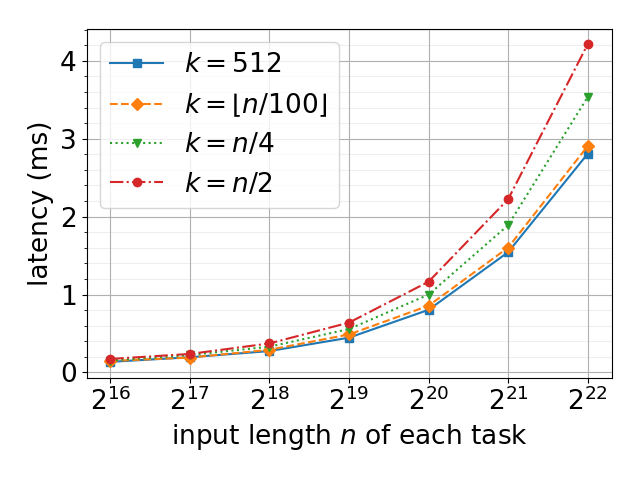}
    \caption{Latency in ms with different $k=O(n)$, batch size $=16$}
    \label{fig:exp:quantile}
\end{figure}

The results are shown in \Cref{fig:exp:quantile}, where the output sorting cost is excluded, because the output length is very large in these cases.
In general, the execution time increases as $k$ increases.
When $k$ is the median, the latency only increases by 50\% compared with the baseline for $n=2^{22}$, while the value of $k$ has already increased 4096 times, implying great scalability with respect to $k$.

% ------------------------------------------------

\subsection{Robustness Results}
\label{sec:eval:concentrated}

To show the impact of adversarial distributions on radix top-k and the effect of our scaling technique against such impact, we design the following experiments.
As analyzed in \Cref{sec:discussion:robust}, we use uniform distributions with small ranges as representative adversarial distributions.
We evaluate the performance with a fixed $k=512$, varying the input length, with and without scaling, on inputs generated from two distributions: (1) all elements sampled from $\text{Uniform}[0.6, 0.7]$, and (2) all elements sampled from $\text{Uniform}[128.6, 128.7]$.
These two distributions differ from each other in that the elements in case (1) can still be distinguished by the most significant 12-bit digit, while all elements in case (2) share the same most significant digit, implying an adversarial input distribution.

\begin{figure}[htb]
    \centering
    \subfigure[$\text{Uniform}{[0.6, 0.7]}$, ours only]{
        \centering
        \includegraphics[width=0.4\linewidth]{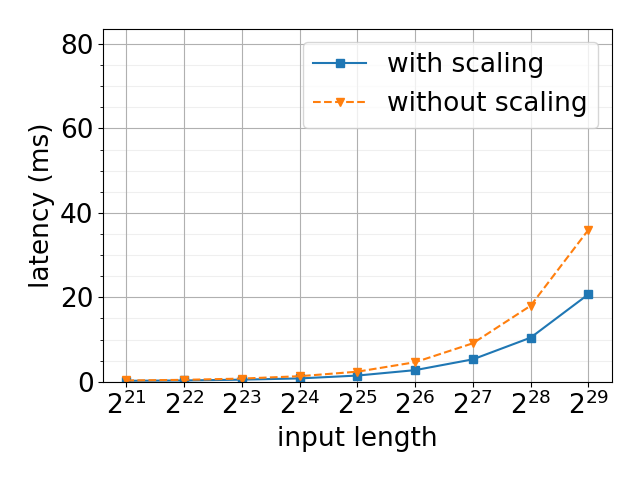}
        \label{fig:exp:scale:good}
    }
    \subfigure[$\text{Uniform}{[128.6, 128.7]}$, ours only]{
        \centering
        \includegraphics[width=0.4\linewidth]{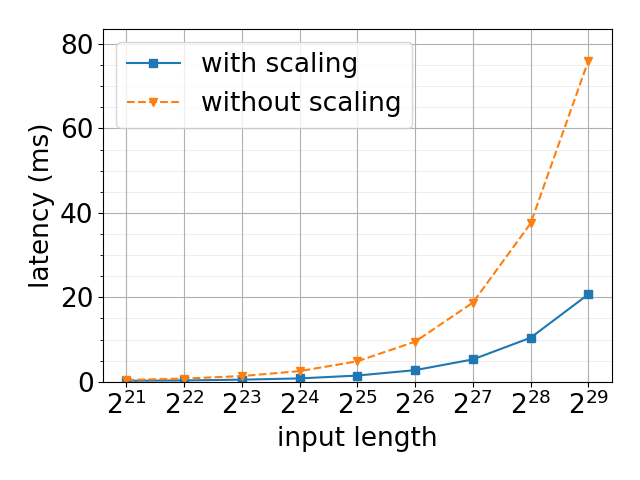}
        \label{fig:exp:scale:bad}
    }
    \\
    \subfigure[$\text{Uniform}{[0.6, 0.7]}$, all]{
        \centering
        \includegraphics[width=0.4\linewidth]{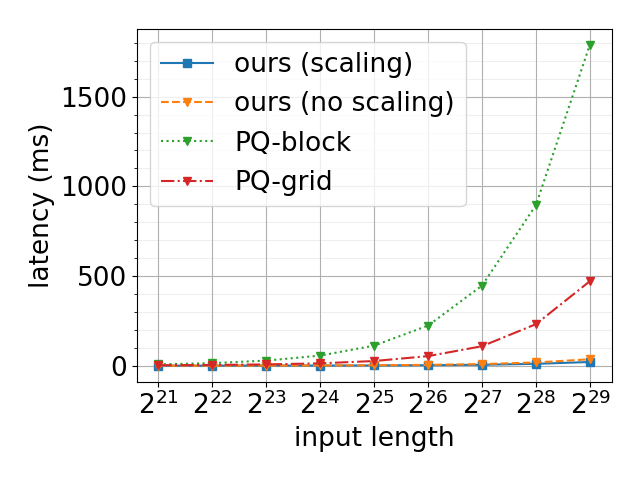}
        \label{fig:exp:scale:good-all}
    }
    \subfigure[$\text{Uniform}{[128.6, 128.7]}$, all]{
        \centering
        \includegraphics[width=0.4\linewidth]{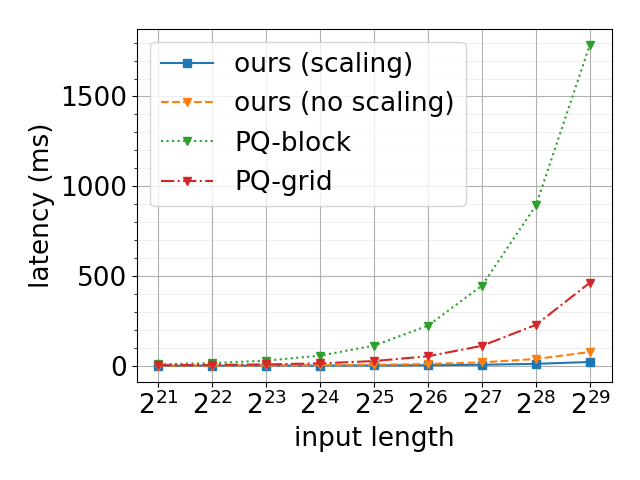}
        \label{fig:exp:scale:bad-all}
    }
    \caption{Latency in ms with and without scaling technique}
    \label{fig:exp:scale}
\end{figure}

The results are shown in \Cref{fig:exp:scale}.
In both cases, scaling helps improve performance against adversarial distributions.
In particular, in case (2) where the most significant digit cannot distinguish any two of the input keys, scaling improves the performance by up to a factor of 2.7.
Besides, \Cref{fig:exp:scale:good-all,fig:exp:scale:bad-all} show that in both cases our method significantly outperforms the prior art, no matter with or without scaling.
The other methods are insensitive to the input distribution, and with scaling, our method is also able to perform as well in case (2) as in case (1).

\begin{figure}[htb]
    \centering
    \subfigure[ours only, without scaling]{
        \centering
        \includegraphics[width=0.4\linewidth]{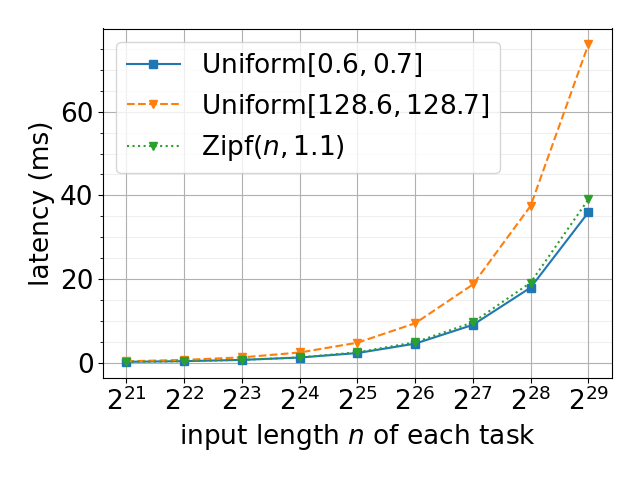}
        \label{fig:exp:zipf:ours}
    }
    \subfigure[comparison with the prior art]{
        \centering
        \includegraphics[width=0.4\linewidth]{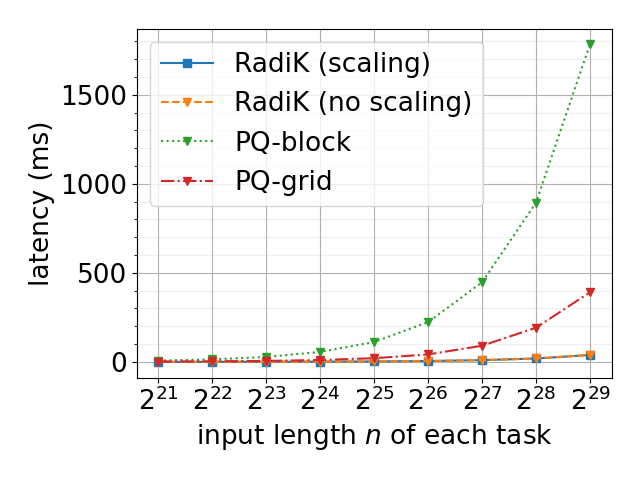}
        \label{fig:exp:zipf:all}
    }
    \caption{Performance comparison on Zipfian distribution}
    \label{fig:exp:zipf}
\end{figure}

According to the analysis in \Cref{sec:discussion}, the cases above are sufficient to illustrate the robustness of our method against adversarial distributions.
However, to make the evaluation more thorough, we also conduct the same test on the Zipfian distribution, a commonly known skewed distribution.
We sample inputs from a Zipfian distribution with skewness equal to 1.1.
\Cref{fig:exp:zipf:ours} shows that the impact of $\text{Zipf}(n,1.1)$ distribution on our method is similar to $\text{Uniform}{[0.6, 0.7]}$, far less adversarial than $\text{Uniform}{[128.6, 128.7]}$.
Moreover, \Cref{fig:exp:zipf:all} shows that our method still significantly outperforms the others with the Zipfian input distribution.

We do not include bitonic in this section of the evaluation.
Note that bitonic is insensitive to data distribution, so its performance in these cases should be the same as the cases with inputs following $\text{Uniform}{[0, 1]}$.
Since our method with scaling performs as well in these cases as it does in \Cref{fig:exp:multibatch}, the comparison between our method with scaling and bitonic should also be the same as \Cref{fig:exp:multibatch}, which does not change the conclusions in this section.

\section{Related Work}
\label{sec:related}

\subsection{Parallel Sorting on GPUs}
\label{sec:related:sort}

Sorting and its derivatives are building blocks of practical GPU top-k algorithms.
A number of GPU sorting algorithms have been studied in recent years~\cite{batcher_sort_1968, hagen_bitonicsort_2010, merrill_radixsort_2011, elias_radixsort_2017,Govindaraju_2006_terasort,satish_2009_sorting, cho_2015_paradis}, among which Bitonic Sort and Radix Sort are two representatives.
Bitonic Sort~\cite{batcher_sort_1968, hagen_bitonicsort_2010} works atop bitonic sequences, each consisting of two subsequences of the same length sorted in opposite directions. 
A bitonic sequence can be sorted in parallel by a sorting network.
The vanilla Bitonic Sort can only deal with inputs of length equal to a power of 2~\cite{batcher_sort_1968}, while a virtual padding technique~\cite{hagen_bitonicsort_2010} extends it for arrays of arbitrary size.
Similar to Radix Select, Radix Sort~\cite{merrill_radixsort_2011, elias_radixsort_2017,satish_2009_sorting} scatters elements into bins, but this time every bin needs to be sorted and the output is the concatenation of all bins.
Global memory access is also a problem, and to improve memory bandwidth, Hybrid Radix Sort performs block-local sort for bins whose sizes are small enough~\cite{elias_radixsort_2017}. 
However, this does not help in Radix Select, because Radix Select uses a prefix sum to find the pivot instead of sorting any bin.

\subsection{Workload Reduction for Top-K Selection}
\label{sec:related:workload}

As mentioned in \Cref{sec:background:topk:radix}, the key to distribution selection is to reduce the problem size by focusing on the bin containing the $k$th element.
There is other work aimed at reducing the top-k workload in general scenarios.
Dr. Top-K~\cite{2021_Gaihre_dr_topk} is a delegate-centric system that helps reduce the workload of GPU top-k including Radix Select, Bucket Select, and Bitonic Select, by dividing the input into sub-ranges and selecting the delegates from them.
However, its benefit decays as $k$ increases, while our proposed method is insensitive to $k$.
External merge sort~\cite{2020_chronis_external_merge_sort} targets even more enormous use cases where $k$ is so large that the output will not fit in the primary memory.
It is guided by histograms and uses a cutoff filter to help maintain a priority queue, thus reducing the size of the input to be sorted or written to secondary storage.
The scopes of these methods are fundamentally different from ours.

\subsection{Concurrent and Recent Work}
\label{sec:related:concurrent}

A concurrent but independent work on GPU top-k selection~\cite{nv_topk_sc23} extends the discussion in an earlier work~\cite{nvidia_topk_2020}, including Grid Select which is one of our baselines in \Cref{sec:eval}.
It proposes AIR Top-K, which also addresses optimizations and robustness of radix top-k, but in a completely different way than our work.
We focus on optimizing individual kernels of radix top-k, while AIR Top-K focuses on kernel fusion. 
As for robustness, we change the input data distribution with the help of our scaling technique, while AIR Top-K changes its buffering strategy depending on the distribution.
Compared with the recent poster on scalable GPU radix top-k~\cite{ppopp24_poster}, this paper adds substantially new insights into data distribution and robustness as in \Cref{sec:discussion}, and includes a detailed introduction to real-world applications as well as more thorough experiments, especially for quantile scenarios and adversarial distributions.

\section{Conclusion}
\label{sec:conclusion}

We present a scalable GPU top-k that is highly efficient for large $k$ and huge input length regardless of batch size. 
Our carefully designed optimization framework consists of two sets of optimizations: the global memory access optimizations with hierarchical atomics and the flush-efficient write buffer, and the batch query optimizations with task rescheduling and on-demand offset padding. 
Experiments show that our method gains up to 2.5$\times$ speedup for non-batch queries and 4.8$\times$ for batch queries.
Moreover, we address the impact of adversarial distributions with a lightweight scaling technique, which improves the performance by up to 2.7$\times$ on highly adversarial distributions. 
Taken together, these results confirm the efficiency, scalability, and robustness of our method.

% \section*{Acknowledgements}
% This document is derived from previous conferences, in particular ISCA 2024 and HPCA 2021.

%%%%%%% -- PAPER CONTENT ENDS -- %%%%%%%%

%%%%%%%%% -- BIB STYLE AND FILE -- %%%%%%%%
\bibliographystyle{IEEEtranS}
\bibliography{bib/ref.bib}
%%%%%%%%%%%%%%%%%%%%%%%%%%%%%%%%%%%%

\end{document}